# Synthesis and characterisation of LK-99

Ivan Timokhin[1,2,+], Chuhongxu Chen[1,2,+], Ziwei Wang[1,2,+], Qian Yang[1,2,*], Artem Mishchenko[1,2,*]

[1]Department of Physics and Astronomy, University of Manchester, Manchester M13 9PL, UK.

[2] National Graphene Institute, University of Manchester, Manchester M13 9PL, UK.

[+] These authors contributed equally

*e-mail: qian.yang@manchester.ac.uk, artem.mishchenko@gmail.com

**Recently, two arXiv preprints (Refs.[1,2]) reported signatures of superconductivity above room temperature (SART) and at ambient pressure, striking worldwide experimental research efforts in replicating the results[3-6], as well as theoretical attempts to explain the purported superconductivity[7-11]. The material of interest has chemical formula $Pb_{10-x}Cu_x(PO_4)_6O$, where $x ≈ 1$, and was named by the authors as LK-99. It belongs to lead apatite family, and was synthesised from two precursors, lanarkite ($PbSO_4·PbO$) and copper phosphide ($Cu_3P$). Here we performed a systematic study on LK-99, starting from solid-state synthesis, followed by characterisation and transport measurements. We did not observe any signatures of superconductivity in our samples of LK-99.**

## Introduction

In the dynamic field of LK-99, in just three weeks after first preprints, over 3 dozen of follow-up works emerge. And this trend seems to continue increasing. Below we briefly review the state of the art of LK-99 research.

Series of DFT calculations[8-15] found ultra-flat electronic bands across all Brillouin zone (plane-flat[16] bands) near the Fermi level in LK-99, suggesting strong electronic correlations. The importance of oxygen for flat bands, as a result from the hybridisation between Cu $d$-states and O $p$-states was highlighted[10,17]. One of preprint even suggested that annealing LK-99 in oxygen atmosphere could improve the band flatness, since O deficiency spoils the orbital hybridisation[17]. Cu-O hybridisation was also discussed in the context of spin-glass-like behaviour of Cu-O hybrids[15]. Overall, however, the band flatness of LK-99 seems to stem from electronic localisation due to a large separation between copper atoms.

Spin-orbit coupling (SOC) could play a role in the electronic structure of LK-99, as DFT calculations without SOC give flat band at the Fermi level, while DFT with SOC results in band gap opening; the conduction band is still very flat (bandwidth 0.025 eV), and could be reached upon electron doping of 0.5 e/unit cell[13]. Besides, SOC could lead to a ferromagnetic ground state[13]. DFT calculations without explicitly including SOC performed in one study revealed antiferromagnetic ground state, their work also shows that LK-99 is a semiconductor with 0.9 eV band gap[18]. Another study, however, amounts SOC to only 3 meV gap in the flat band of LK-99, making it less important comparing to other effects such as exchange interactions (e.g., antiferromagnetic superexchange)[19].

The importance of copper nanoclusters for SART is emphasized in one study, based on the author's previous development of resonating valence bond (RVB) theory, so-called atom Mott insulator (AMI)[7]. It suggests that plane-flat bands of LK-99 are not that important for SART, and could only result in very low transition temperatures[7]. The low-temperature scale of the corelations in LK-99 is also suggested in local density approximation band calculations[20].

Using tight-binding (TB) calculations, analysis of Berry curvature in flat bands of LK-99 revealed no signatures of magnetism[21]. Another TB study of the plane-flat band in LK-99 showed its topological origin, however no discussions on possible superconductivity or magnetism were provided[22].

A combination of dynamic mean field theory (DMFT) and four- and two-band TB calculations predicts that LK-99 is a Mott insulator with strong interactions at the flat band[23]. A combination of DFT and DMFT calculations also suggested the importance of oxygen for the electronic properties of LK-99, and vast complexity of its electronic structure[24]. The DFT + DMFT approach was further elaborated, together with an update on tight-binding description of LK-99[25]. Meanwhile, the crystal structure of a parent material of LK-99, the undoped lead apatite also known as oxypyromorphite, $Pb_{10}(PO_4)_6O$, has also been recently revisited[26] following the ongoing development of LK-99 research, which should provide better input for future *ab initio* calculations.

Overall, the importance of band flatness and the contribution of copper or oxygen in the electronic structure of LK-99 was highlighted throughout theoretical attempts. Yet more theoretical works to unify the currently dispersed findings are needed to provide a general picture to understand experimental observations on LK-99.

In contrast to a flurry of theoretical responses, experimental attempts to reproduce LK-99 synthesis and test for superconductivity are more sporadic. One early attempt to reproduce LK-99 claimed successful synthesis, supported by X-ray diffractometry (XRD) analysis of samples. However, no signs of superconductivity was observed[27]. They reported weak diamagnetic response of the samples, -0.006 emu/g at 10 kOe at room temperature. Another attempt claims observation of strong diamagnetic behaviour of synthesised LK-99, however no quantitative data were provided beyond the observation that some LK-99 flakes deviate to a large angle under magnetic field of a permanent magnet[6]. Another study reported that synthesised LK-99 samples are heterogeneous, with a small fraction of 'flaky' fragments exhibiting 'half-levitation' over permanent magnet at room temperature[28]. Magnetisation measurements of these fragments (and even of the samples that did not half-levitate) show magnetic response, e.g., 0.00055 emu/g at 10 Oe, but -0.01 emu/g at 10 kOe at room temperature. Low magnetic field data show ferromagnetic hysteresis. Ferromagnetic signatures were even stronger in half-levitating fragments. However, quantitative measurements were not possible due to their small size.

In one preprint, where the authors claimed the observation of superconducting transition at approximately 100 K, with resistance drop following decreasing temperature in LK-99[3]. This result, however, could also be interpreted as a metallic behaviour of the sample, since the noise level of their measurements precludes detecting resistance lower than $10^{-5}$ Ω. Another preprint reported semiconducting behaviour (with resistivity of $10^4$ Ω·cm at room temperature) of LK-99 and no signs of diamagnetic levitation[5].

Further attempts to reproduce LK-99 also show sample inhomogeneity and semiconducting temperature dependence[29]. Curiously, LK-99 shows a small drop in resistance near 387 K (and a small diamagnetic jump at temperatures in the 340-350 K range), without any signs of zero resistance at any studied temperatures. The authors attributed resistance drop to either LK-99 itself, or impurities like $Cu_2S$ or $Cu_2O$. The presence of $Cu_2S$, and its crucial role in the behaviour of LK-99 was elaborated in another experimental work[30]. By measuring resistance of pure $Cu_2S$ and LK-99 containing $Cu_2S$, the authors reproduced the sudden drop in resistance around 385 K, attributing it to a structural phase transition of $Cu_2S$ from β to γ phase, accompanied by 3-4 orders of magnitude decrease in resistivity. The role of $Cu_2S$ impurities in explaining LK-99 physics was also highlighted elsewhere[31], signifying the importance of obtaining $Cu_2S$-free LK-99 for further investigations. LK-99 is also found to be hygroscopic and could convert from $Pb_9Cu_1(PO_4)_6O$ to $Pb_9Cu_1(PO_4)_6(OH)_2$ upon expose to ambient moisture[32]. Yet, again, no signs of superconductivity (or theoretical hints towards it) were observed.

So far, none of the experimental efforts has successfully reproduced SART in LK-99 as reported in the original preprints. Rather, the reported results show a range of diverse behaviour that likely originated from the inhomogeneity of the synthetic LK-99 materials, because of a lack of precise recipe. The ongoing enthusiasm in searching for SART calls for collective efforts in open research to establish unified synthesis routes and characterization standard, for LK-99, and for future quests towards SART.

## Results and discussion

In our work, several attempts to synthesize LK-99 were carried out. The main synthesis route, described in the arXiv preprints[1,2] and in the patent[33], rely on the reaction between lanarkite ($PbSO_4·PbO$), and copper phosphide ($Cu_3P$) at 925°C in a sealed environment. Our synthesis follows closely the original preprints and patent, the details are described below. We have also tried different ratios of precursors, various container/crucible materials, and a range of temperature profiles.

## Precursors synthesis

In our experiments, lanarkite was first synthesized following the procedure described elsewhere[1,2,33]. Briefly, 1:1 molar ratio of PbO and $PbSO_4$ (first synthesis was carried out with compounds prepared from lead nitrate: PbO - by decomposition at 600°C, $PbSO_4$ – by precipitation with $H_2SO_4$; further attempts were made from reagents purchased from commercial sources) were ground in agate mortar to form a fine mixture. Then the mixture was heated at 750°C in an alumina crucible for 24 hours in air. The mixture was left to cool naturally before taking out of the furnace. X-ray diffractometry (XRD, Rigaku Miniflex I) was used to confirm the crystalline structure of the as-synthesized material. Its XRD peaks (Fig. 1a) match well with the lanarkite reference entry, confirming its purity.

We then synthesized copper phosphide ($Cu_3P$) according to the procedure described before[1,2,33]. To this end, a stoichiometric ratio of copper powder (10 µm, 99.95%) and red phosphorus (Chempur 99.995%) were mixed together in the argon glovebox and sealed in the quartz ampule under $10^{-3}$ mbar vacuum. The ampule was then heated to 550°C and kept for 15 hours, then at 650°C for another 10 hours. After cooling down naturally, we opened the ampoule and took the sample out for XRD measurements. XRD results (Fig. 1b) confirm the obtained compound as $Cu_3P$ and its good crystallinity.

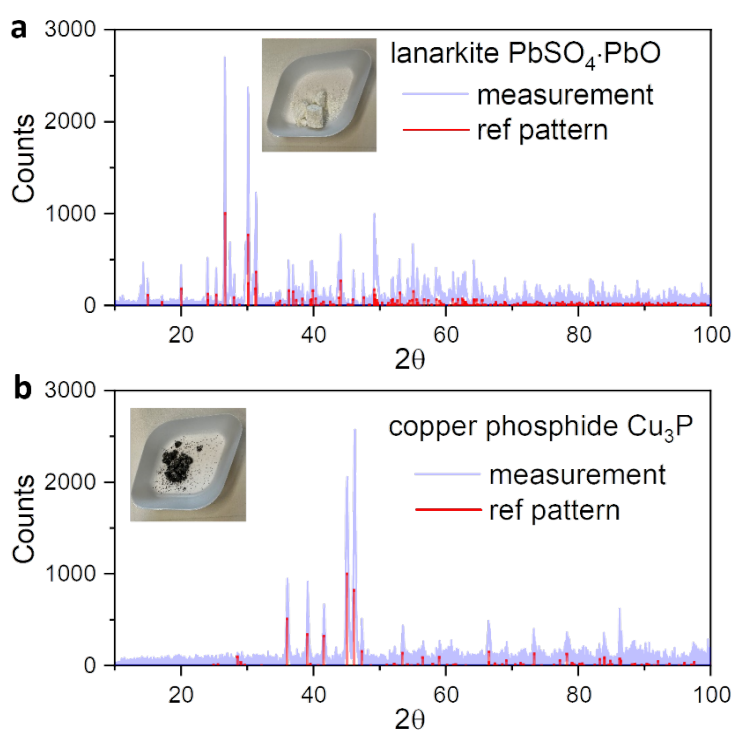

**Fig. 1. XRD patterns of synthesized precursors. a,** lanarkite. **b,** copper phosphide. Insets show photographs of the materials.

With the two precursors ready, we can start the synthesis of LK-99. To make sure a good mixture of the two components, we weigh and grind both components into fine powder before initiating the reaction. $Cu_3P$ is mechanically hard and brittle, it was ground in agate mortar inside glovebox and sieved through a 300-mesh sieve. Lanarkite was ground at ambient environment. Two precursors were mixed and sealed inside quartz ampoule and heated up for certain period of time. Afterwards, samples were taken out of the furnace for further characterization and measurements. We found that molten lead compounds severely corrode quartz glass, leading to the cracking of the quartz ampoule. The obtained materials were first characterized using XRD, scanning electron microscopy (SEM), and energy-dispersive X-ray (EDX) microanalysis (Zeiss Ultra microscope), to confirm the crystallinity and elemental composition of the synthesized materials.

## Copper-doped lead apatite (LK-99) synthesis and characterisation

Lanarkite and $Cu_3P$ were mixed with compound molar ratio 1:1, as described previously, and loaded directly into quartz ampoules. The ampoules were heated to 925°C in a furnace and kept for 15 hours. After the ampoules cooled down, we found severe cracking or even the total breakdown of the ampoules. To prevent this from happening, we then tried loading the precursor mixture into alumina crucibles, before loading into quartz ampoules for high-temperature reaction. This helped to some extent, to contain the reaction and to reduce the damage to the ampoules from molten reactants. Figure 2a shows one of the quartz ampoules after reaction at high temperature, similar to that reported in Fig. 1f in the original 6-author preprint[2]. However, alumina crucible also got heavily corroded by lead-containing melt during growth (Fig. 2e,f). EDX of the as-grown sample shows lead-containing matrix with islands of aggregated metallic copper (Fig. 2b,c,d). The obtained sample was also checked under XRD, to compare with the original LK-99 sample[2]. The XRD peaks of the synthesized material match only partially with that of undoped lead apatite, Fig. 2g.

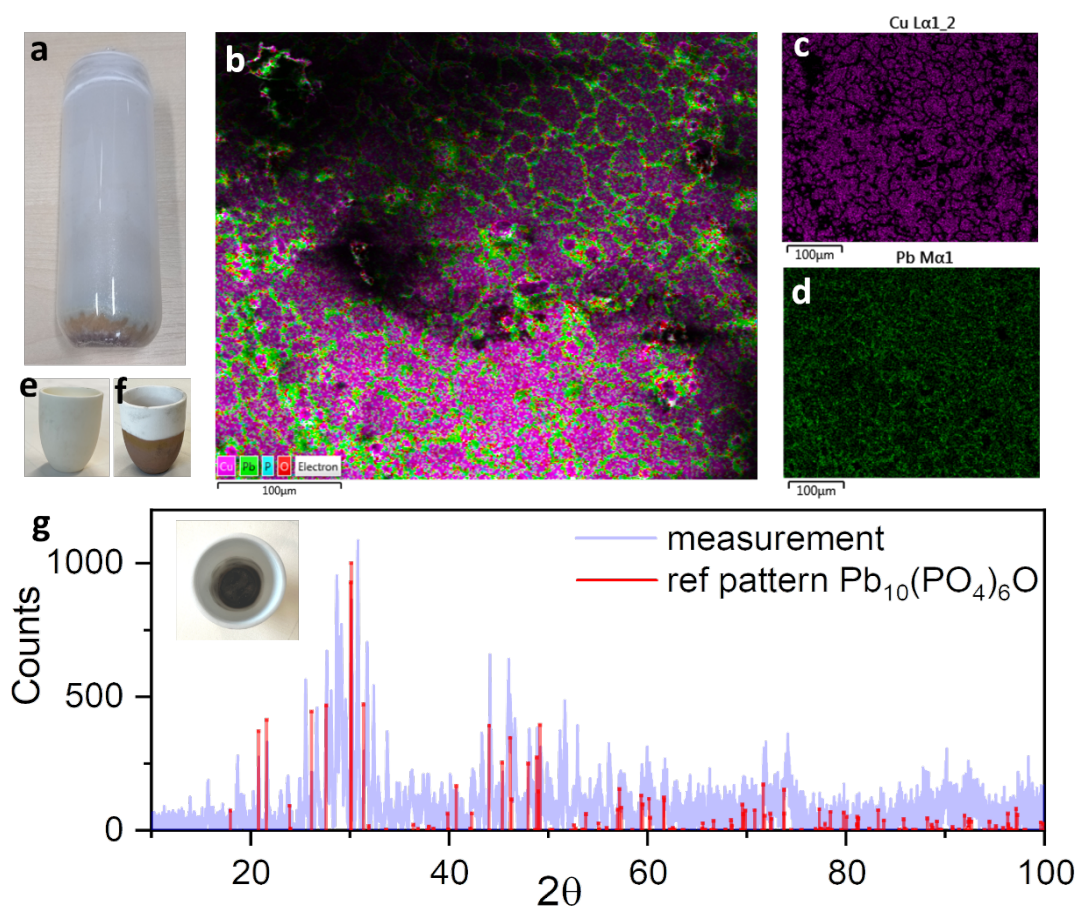

**Fig. 2. Synthesis and characterisation of LK-99. a,** photograph of the ampoule after synthesis showing damage due to volatile lead compounds. **b,** SEM micrograph with overlay of EDX map showing elemental composition of the synthesised LK-99. **c,d,** elemental maps of Cu and Pb of the same specimen as in panel b. The EDX mapping reveals segregation of metallic copper islands in the matrix of lead apatite. **e,f,** photographs of alumina crucibles before (e) and after (f) synthesis. Lead-containing melt causes strong corrosion of alumina as evidenced by the coloration. **g,** XRD of the synthesized material and its comparison with the reference spectrum of lead apatite. Inset shows photograph of as-grown LK-99 inside alumina crucible.

We then used copper crucible, in replacement of alumina crucible, to repeat the experiment, in order to avoid crucible corrosion and the potential Al contamination in the synthesized products. Similarly, the synthesis process follows that described previously. XRD and EDX were used to confirm the composition of the final products, Fig. 3. EDX confirms that the final product is made of mainly four elements (Pb, Cu, P and O). Among them, copper tends to form aggregates (Fig. 3b) after precipitating at high temperature (see also Fig. 2b,c). All other elements show more uniform distribution throughout the measured area (more than 3 areas are checked using EDX, showing similar overall distribution). XRD confirmed the presence of LK-99 (Cu-doped $Pb_{10}(PO_4)_6O$), Fig. 3f. Our experimental peaks match what is expected for Cu-doped lead apatite, with the addition of a peak at around 27 degree, as reported in the 6-author preprint[2].

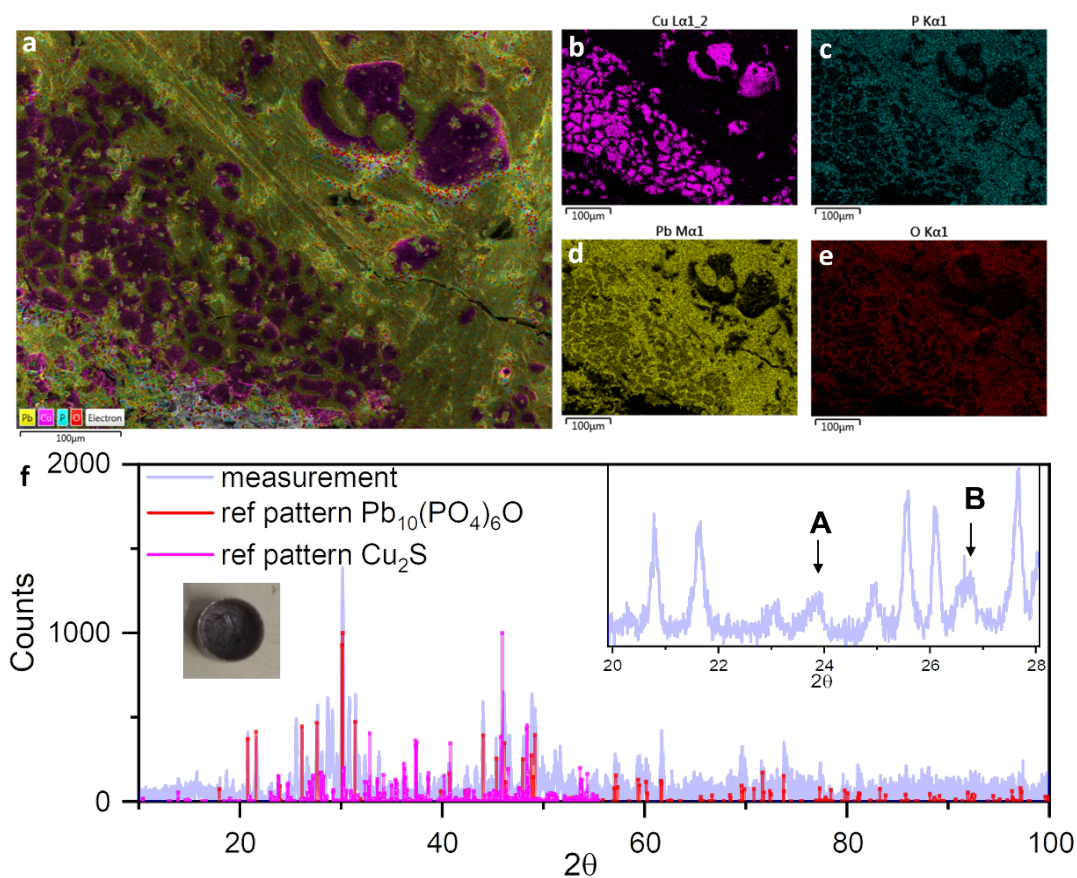

**Fig.3. Synthesis of LK-99 in copper crucible. a-e,** SEM and EDX analysis of as-grown LK-99. **f,** XRD of LK-99. Left inset shows LK-99 inside copper crucible. Right inset shows zoomed-in region of panel f, highlighting peaks that are characteristic of Cu-doped lead apatite, according to Ref. 2.

During the synthesis in copper crucible, some material evaporated and condensed at the bottom of the quartz ampoule. We also characterized this material. It appears to be lead apatite ($Pb_{10}(PO_4)_6$) (see XRD in Supplementary Fig. 1d). EDX results of the material from the bottom wall of the ampoule show that Cu is also present, Supplementary Fig. 1.

To avoid evaporation of the material from copper crucible, we further modified the synthesis procedure. To this end, we covered the copper crucible with copper cap and filled the ampoule with 0.15 bar of argon for the reaction to take place. Temperature and duration of the reaction were kept the same as above, 925°C for 15 hours. This modification completely prevented evaporation of material outside the crucible and helped to contain reaction completely within the crucible. This also prevented side reactions of precursors or LK-99 with quartz ampoule. The obtained LK-99 was still rather inhomogeneous, Fig. 4.

We did not observe either diamagnetic levitation or zero resistance in any of our LK-99 samples. However, in some of the synthesized samples, we found few very small pieces of materials, either black or pink appearance under optical microscope, with strong para- or ferromagnetic response. That is, these samples move, flip, or stand as the magnet moves underneath, similar to those reported in some videos (e.g., as in Ref. 6) in an attempt to replicate LK-99 diamagnetic levitation. We carefully separated these particles and performed EDX analysis on them. As confirmed by EDX, these magnet-responsive particles have iron inclusions (Supplementary Fig. 2). This helps to explain our results with magnetic response. Unfortunately, despite all the care has been taken to avoid iron contamination during the process, some remains.

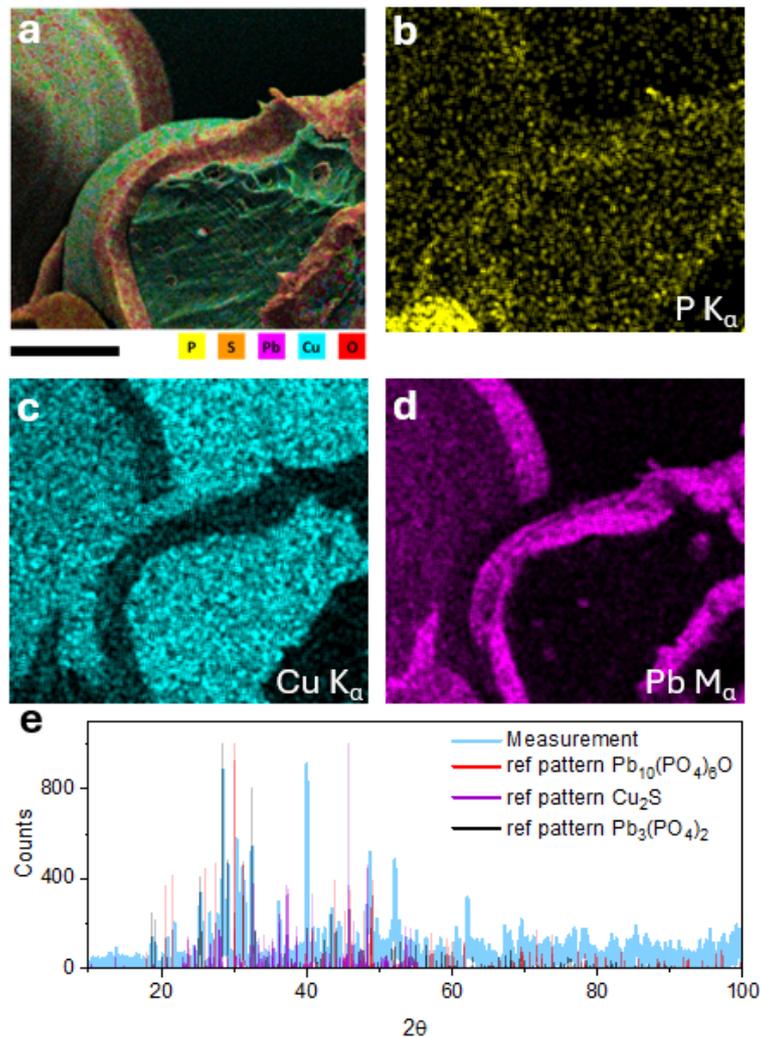

**Fig.4. Synthesis of LK-99 in copper enclosure under argon. a,** SEM image with overlay of EDX map showing elemental composition, **b-d,** EDX analysis for phosphor, copper and lead, respectively. **e,** XRD spectrum of the synthesised LK-99.

## Transport measurements

We then fabricated several samples for transport measurements, both the as-grown material, and a ground and pressed pellet from the synthesized LK-99, details are shown in Fig. 5 and Supplementary Fig. 3. Four-probe resistance was measured using lock-in amplifier, excitation current was in the mA range, and was monitored by a second lock-in. Current-voltage characteristics were recorded using 2614b Keithley SourceMeter and 2182A Keithley Nanovoltmeter, using either 4-terminal or van der Pauw geometry. Temperature of the sample mounted in the vacuum chamber was controlled using Ricor K535LV cryocooler. The temperature was measured by Cryo-Con Model 12 temperature monitor equipped with Pt100 RTD thermometer attached to the sample holder. Due to strong inhomogeneity of the samples, their voltage-current characteristics show a broad variation of response, with resistance ranging many orders of magnitude, and V-I curves showing signatures of switching, hysteresis, nonlinearity, and even negative differential resistance. Samples appear to be brittle as well. Often, samples crack and break down during measurements, when high current (> 50 mA) was applied, or during mounting due to even slight mechanical vibration.

Figure 5c shows temperature dependence of the resistance of one of our samples. High-temperature data show clear drop in resistance, very similar to the one shown in Fig. 5 in the original preprint[2]. This behaviour is very similar to resistance drop in $Cu_2S$ due to structural phase transition, see data in Supplementary Fig. 4 and similar results in Ref. 30. Resistance peaks at 100-150 K are measurement artifacts due to divergence of contact resistance at low temperatures, see the strong increase in out-of-phase signal of measured voltage in the corresponding low temperature range in the inset of Fig. 5c.

Overall, the device shows a combination of resistance drop at high temperature, most likely due to structural phase transition of $Cu_2S$, combined with low-temperature semiconducting behaviour of LK-99.

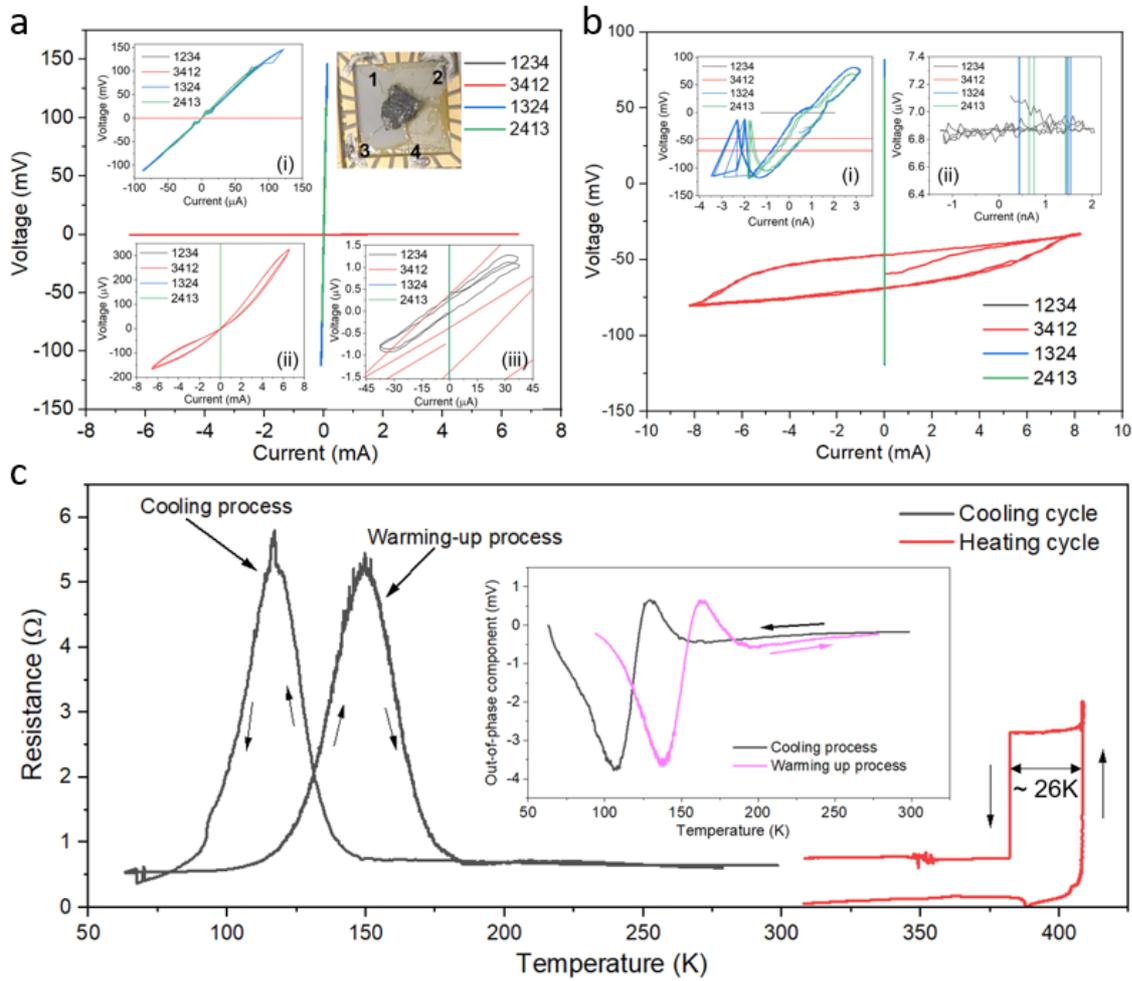

**Fig. 5. Transport measurements of LK-99. a,** van der Pauw voltage-current measurements of as-grown LK-99 sample at room temperature in air. Top-right inset shows a photograph of the device with four terminals marked with numbers. Four numbers in the legend are indications of the measurement configurations: terminals corresponding to the first two numbers are connected to a current source, while terminals corresponding to the last two numbers are connected to a nanovoltmeter. Insets (i), (ii), and (iii) are zoomed-in V-I curves for different measurement configurations. **b,** van der Pauw voltage-current measurement of the LK-99 sample shown in **a** at 62 K in vacuum. The legends and insets (i) and (ii) in **b** are also zoomed-in curves for different measurement configurations. Low-temperature V-I curves show that parts of the device become insulating, most likely due to a granular nature of LK-99 sample. **c,** Temperature dependence of resistance of the LK-99 sample shown in **a**. The heating cycle is first conducted in air and the resistance of the sample is recorded through via dc measurements. Following the heating cycle, the cooling cycle is carried out under vacuum conditions. The transport properties of the sample in the cooling cycle are assessed utilizing lock-in measurements. Inset: the out-of-phase component of the lock-in signal during the cooling cycle. Arrows indicate the temperature change directions.

In summary, we can conclude that although the synthesis of LK-99 shows XRD features similar to the reported, and EDX data confirm their elemental composition, no signatures of superconductivity are visible. We did observe a drop in the resistance of some of our samples at 380-390 K, which is consistent with structural phase transition of $Cu_2S$ in that temperature range.

## Acknowledgements

This research was supported by the European Research Council (ERC) under the European Union's Horizon 2020 research and innovation program (Grant Agreement No. 865590) and the Research Council UK (BB/X003736/1). Q.Y. acknowledges the Royal Society University Research Fellowship URF\R1\221096, and the Research Council UK (EP/X017575/1).

## Data availability

Relevant data are available from the corresponding authors on request.

# Supplementary Information

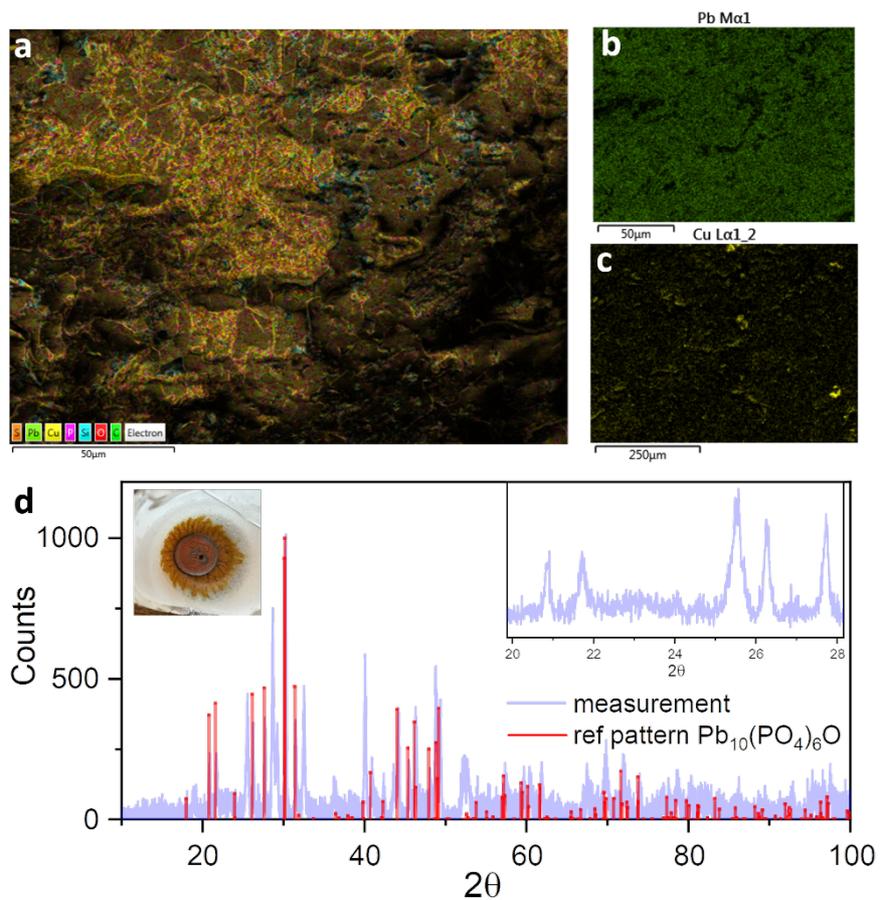

**Supplementary Fig. 1. Lead apatite condensed outside copper crucible in the quartz ampoule. a-c,** SEM and EDX analysis of Cu-doped lead apatite. **d,** XRD of the evaporated material. Left inset shows a photograph of the precipitate at the bottom of quartz ampoule that was used for analysis.

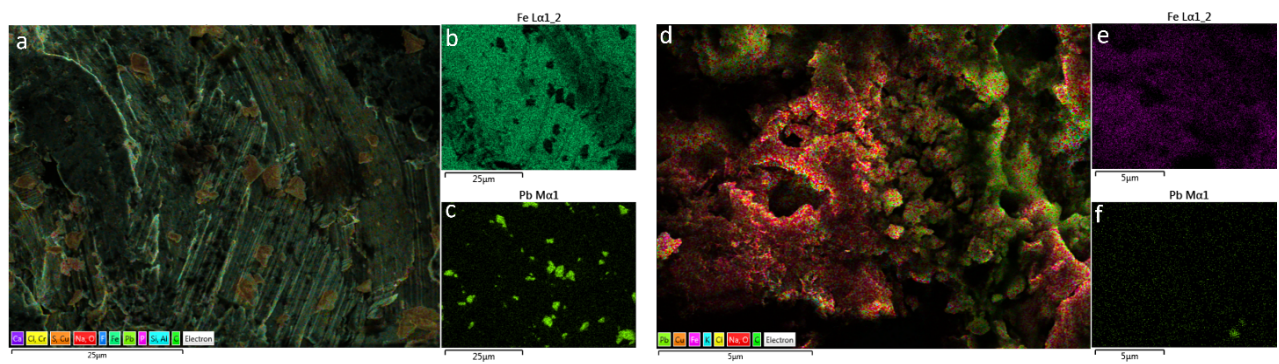

**Supplementary Fig. 2. SEM and EDX analysis of two samples (a-c and d-f) with magnetic response, showing presence of iron.**

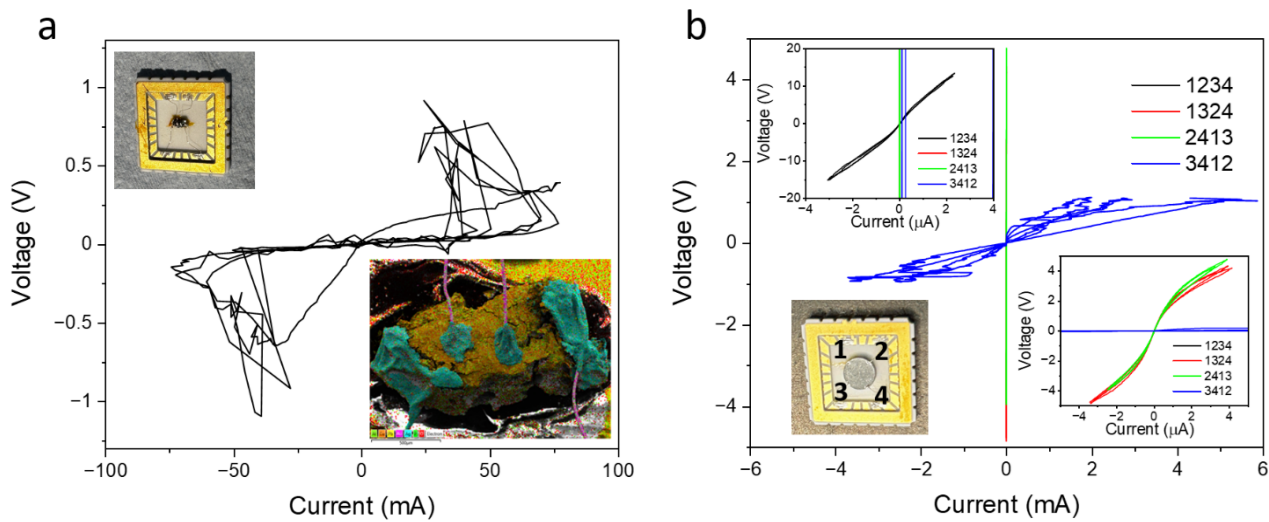

**Supplementary Fig. 3. Transport measurements of LK-99 at room temperature in air. a,** 4-terminal measurement of as-grown LK-99 sample. V/I curve is highly unstable. Left inset shows photograph of the measured sample, right inset is SEM image of the same sample overlayed with EDX map after measurements, cracks formed during measurement is visible in the middle of the sample. **b,** van der Pauw measurements of ground LK-99 pressed into a pellet. Four numbers 1234 show measurement configuration, +I-I+V-V. Insets: top left and right bottom show zoomed-in curves highlighting huge differences in resistance across the pellet. Bottom left inset shows photograph of the pellet with contacts labelled.

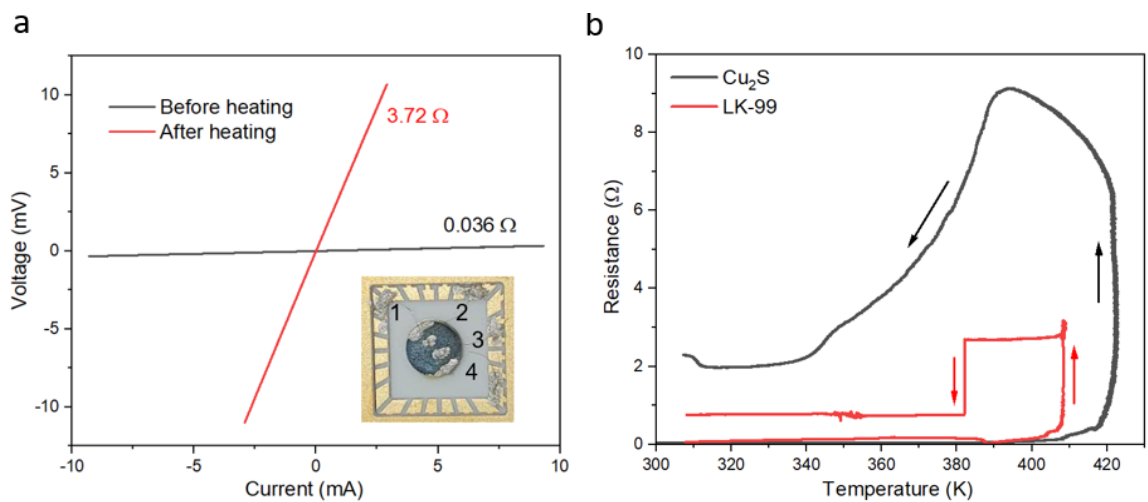

**Supplementary Fig. 4. Transport measurements of $Cu_2S$ pellet. a,** four-terminal measurement of a $Cu_2S$ pellet pressed from grounded $Cu_2S$ powder. Inset: photograph of the pellet with four electrical contacts marked with numbers. A current source is applied through terminals 1 and 4, and terminals 2 and 3 are connected to the high and low of a voltmeter. Two voltage-current curves are measured before and after the heating process shown in **b**, and the corresponding resistances are calculated from the linear fit of the curve. **b,** resistance of the $Cu_2S$ pellet and the LK-99 sample (same data as in Fig. 5c in the main text) at high temperature in the air. The resistance of the sample is recorded via dc measurements. Arrows indicate directions of temperature change.